\documentclass{article}
\usepackage{dcase2019,amsmath,graphicx,url,times,booktabs, tabularx,amssymb,multirow}

\title{TRAINING SOUND EVENT DETECTION ON A HETEROGENEOUS DATASET}

\name{Nicolas Turpault\thanks{This work was made with the support of the French National Research Agency, in the framework of the  project LEAUDS “Learning to under-stand audio scenes” (ANR-18-CE23-0020) and the French region Grand-Est. Experiments presented in this paper were carried out using the Grid5000 testbed, supported by a scientific interest group hosted by Inria and including CNRS, RENATER and several Universities as well as other organizations (see https://www.grid5000).}, Romain Serizel
    }

\address{Universit{\'e} de Lorraine, CNRS, Inria, Loria, F-54000 Nancy, France}

\begin{document}

\ninept
\maketitle

\begin{sloppy}

\begin{abstract}
Training a sound event detection algorithm on a heterogeneous dataset including both recorded and synthetic soundscapes that can have various labeling granularity is a non-trivial task that can lead to systems requiring several technical choices. These technical choices are often passed from one system to another without being questioned. We propose to perform a detailed analysis of DCASE 2020 task 4 sound event detection baseline with regards to several aspects such as the type of data used for training, the parameters of the mean-teacher or the transformations applied while generating the synthetic soundscapes. Some of the parameters that are usually used as default are shown to be sub-optimal.
\end{abstract}

\begin{keywords}
Sound event detection, weakly labeled data, semi-supervised learning, synthetic soundscapes, ablation study
\end{keywords}

\section{Introduction}
\label{sec:intro}
Ambient sound analysis aims at extracting information from the soundscapes that constantly surround us~~\cite{virtanen2018computational}. Many ambient sound analysis applications, ranging from urban planning to home assisted living, have surfaced within the past first years~\cite{Bello:SONYC:CACM:18,radhakrishnan2005audio,serizel2016,jin2012event,debes2016monitoring, zigel2009method}. Most of these were inspired by the fact that we, humans constantly rely on the soundscapes around us to decide on how to act or react.

Sound event detection (SED) is a task of the ambient sound analysis that consists not only in predicting what sound event did occur in a recording but also to detect when did it happen. Intuitively, the simplest way to solve this problem with a system relying on supervised training would be to use a training dataset composed of so-called strongly labeled soundscapes (with onset and offset timestamps). However strongly labeling a sufficiently large dataset is prohibitive. Strong label annotations are also very likely to contain human errors/disagreement given the ambiguity in the perception of some sound event onsets and offsets. One alternative is too rely on so-called weakly labeled soundscapes (without timestamp) that are considerably cheaper to obtain~\cite{serizel2018_DCASE}. In the case of weakly labeled data, we only have information about whether an event is present in a recording or not. We have no information about how many times the event occurs nor the temporal locations of the occurrences within the audio clip. However, there are quite a few shortcoming in exploiting weakly labels soundscapes~\cite{shah_closer_2018,turpault:hal-02467401,McFee:AutoPool:TASLP:18}. Another, cheaper, option is to generate realistic soundscapes that can then easily be strongly labeled~\cite{salamon2017scaper}. The problem in this latter case is that is that there might be some mismatch between the synthetic soundscapes generated for training and the recorded soundscapes fed to the SED at runtime.

In DCASE 2019 task 4~\cite{turpault_2019} we proposed to try solving the above problems by designing DESED, a dataset composed of weakly-labeled and unlabeled recorded soundscapes and strongly labeled synthetic soundscapes generated with Scaper~\cite{salamon2017scaper}. Exploiting such a heterogeneous dataset all together is not necessarily trivial. The baseline was inspired by previous submission to the challenge~\cite{Lu2018,Delphin-Poulat2019} and developed incrementally (and so are quite a few systems trying to solve this task). These two aspects led to a solution that is involving a lot of different technical choices that were rarely clearly motivated in the literature. These choices are frequently passed from one system to another without being questioned.

The aim of this paper is to propose a detailed analysis of DCASE 2019 baseline~\cite{turpault_2019}. The ablation study includes the analysis of aspects such as the kind of data used for training, the parameters for the mean-teacher or the transformation applied on the synthetic soundscapes dataset. The conclusion of this study leads to a system that is close to the baseline for DCASE 2020 task 4.
\section{Problem statement and baseline description}
\begin{figure*}
 \includegraphics[width=\textwidth]{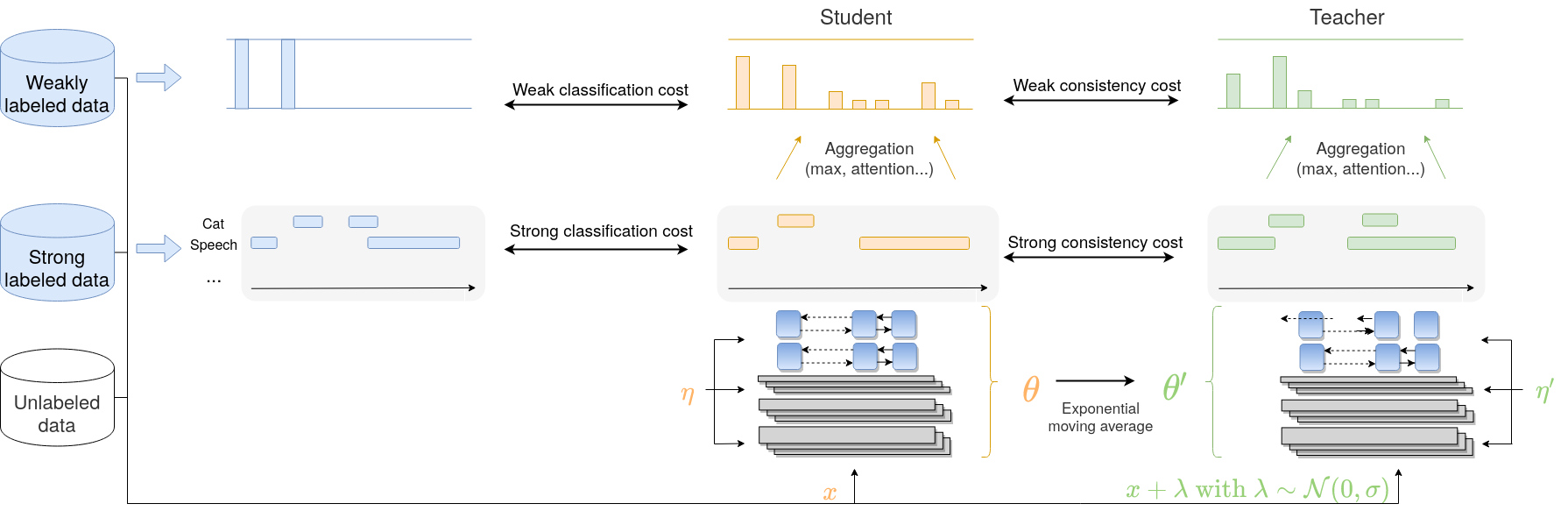}
 \caption{Mean-teacher model. $\eta$ and $\eta'$ represent noise applied to the different models (in this case dropout).}
\label{fig:sed_baseline}
\end{figure*}
\label{sec:pb}
\subsection{Problem description}

The problem we try to solve here is similar to that of DCASE 2019 Task 4~\cite{turpault_2019} and focuses on the same 10 classes of sound events. Systems are expected to produce strongly labeled output (i.e.~detect sound events with a start time, end time, and sound class label), but are provided with weakly labeled data (i.e.~sound recordings with only the presence/absence of a sound event included in the labels without any timing information) for training.
Multiple events can be present in each audio recording, including overlapping events. As in the previous iteration of this task, the challenge entails exploiting a large amount of unbalanced and unlabeled training data together with a small weakly annotated training set to improve system performance and an additional training set with strongly annotated synthetic soundscapes which can be created in many ways.

\subsection{Sound event detection baseline}
\label{sub:sed_baseline}

The SED baseline system is inspired by the best performing system from DCASE 2018 Task 4~\cite{Lu2018} and the improvement provided by the second best performing system in DCASE 2019 task 4~\cite{Delphin-Poulat2019} . It uses a mean-teacher model which is a combination of two models: a student model and a teacher model (both have the same architecture). The student model is the final model used at inference time, while the teacher model aims at helping the student model during training and its weights are an exponential moving average of the student model's weights. A depiction of the baseline model is provided in Figure \ref{fig:sed_baseline}.

Our implementation of the mean-teacher model is based on the work of Tarvainen and Valpola~\cite{tarvainen_mean_2017}. The models are a combination of convolutional neural network (CNN) and recurrent neural network (RNN) called CRNN. The model architecture is inspired by DCASE 2019 task4 second to best system.

The student model is trained on the strongly and weakly labeled data. The loss (binary cross-entropy) is computed at the frame level for the strongly labeled synthetic data and at the clip level for the weakly labeled data. The teacher model is not trained, rather, its weights are a moving average of the student model (at each epoch). During training, the teacher model receives the same input as the student model but with added Gaussian noise, and helps train the student model via a consistency loss (mean-squared error) for both strong (frame-level) and weak (clip-level) predictions for all the data in the batch. Every batch contains a combination of unlabeled, weakly and strongly labeled samples.

\begin{table*}
\centering
\begin{tabular}{|l|c|c|c|c|c|c|}
\hline
Training set&\multicolumn{6}{c|}{Ratio between training sets}\\
\hline
Synthetic & 1/3 & 1 & 1/4 &  & & 1/2 \\
Weak &1/3 &&&1&1/4&1/2 \\
Unlabeled&1/3 &&3/4&&3/4&\\
\hline
F1-score & 34.14\%&	20.41\%&11.56\%&	16.46\%&	17.97\%&	31.76\% \\
PSDS & 0.502 & 0.250&0.140&	0.287&	0.328& 0.435 \\
\hline
\end{tabular}
\caption{SED performance on the evaluation set depending on the kind of data used for training.}
\label{tab:train_set}
\end{table*}

\begin{table}
\centering
\begin{tabular}{|l|c|c|c|}
\hline
\multicolumn{1}{|c|}{}&\multicolumn{3}{c|}{Pitch-shifting} \\
\multicolumn{1}{|c|}{Training}&  &  &  \checkmark\\
\multicolumn{1}{|c|}{x-valid}&  & \checkmark & \checkmark \\ \hline
F1-score &35.15\%&35.91\%&34.14\% \\
PSDS & 0.487&	0.495&	0.502\\
\hline
\end{tabular}
\caption{SED performance depending on pitch shifting being applied or not on isolated events during synthetic soundscapes generation.}
\label{tab:pitch_synth}
\end{table}

\begin{table}
\centering
\begin{tabular}{|l|c|c|c|}
\hline
\multicolumn{1}{|c|}{}&\multicolumn{3}{c|}{Reverberation} \\
\multicolumn{1}{|c|}{Training}&\checkmark  &  &  \\
\multicolumn{1}{|c|}{x-valid}&\checkmark  & \checkmark &  \\ \hline
F1-score &18.30\%&	35.54\%&34.14\% \\
PSDS & 0.435&	0.508&	0.502\\
\hline
\end{tabular}
\caption{SED performance depending on reverberation being applied or not on during synthetic soundscapes generation.}
\label{tab:reverb_synth}
\end{table}
\begin{table}[t!]
\centering
\begin{tabular}{|l|c|c|c|c|}
\cline{2-5}
\multicolumn{1}{c|}{}&\multicolumn{4}{c|}{Mean-teacher SNR} \\
\multicolumn{1}{c|}{}& 0 & 15 & 30 & $\infty$\\ \hline
F1-score &12.56\%&26.29\%&34.14\%&37.80\%\\
PSDS & 0.261	&0.437	&	0.502&0.540\\
\hline
\end{tabular}
\caption{SED performance depending on the SNR parameter for the noise applied to the mean-teacher input.}
\label{tab:SNR}
\end{table}

\begin{table*}
\centering
\begin{tabular}{|l|c|c|c|c|c|c|}\hline
Ramp-up CC&&&\checkmark&&&\checkmark\\
Ramp-up LR&&&&\checkmark&\checkmark&\checkmark\\
Consistency weight&1&2&[0, 2]&1&2&[0, 2]\\
\hline
F1-score	&24.20\%	&27.96\%	&25.28\%	&31.97\%	&33.05\%	&34.14\%	\\
PSDS	&0.408	&0.412	&0.420	&0.489	&0.475	&0.502	\\
\hline
\end{tabular}
\caption{Impact Of the ramp-up}\label{tab:rampup}
\end{table*}

\begin{table}
\centering
\begin{tabular}{|l|l|c|c|c|c|}
\hline
\multicolumn{2}{|c|}{Reverb}&\checkmark  & \checkmark & & \\
\multicolumn{2}{|c|}{Pitch-shifting}&\checkmark  & & \checkmark & \\ \hline
\multirow{2}{*}{VAL}&F1-score &31.13\%&	36.27\%&37.80\%& 34.46\%\\
&PSDS & 0.482&	0.521&	0.540&0.520\\
\hline
\multirow{2}{*}{EVAL}&F1-score &33.7\%&	39.9\%&39.0\%& 36.8\%\\
&PSDS & 0.515&	0.568&	0.552&0.566\\
\hline
\end{tabular}
\caption{Parameter combination for the new baseline definition}
\label{tab:reverb_pithc_synth}
\end{table}

\section{Baseline setup and dataset}
\label{sec:base_bdd}
\subsection{DESED dataset}
\label{subsec:desed}
The dataset used for SED experiments is DESED\footnote{\url{https://project.inria.fr/desed/}}, a flexible dataset for SED composed of 10-sec audio clips recorded in a domestic environment or synthesized to simulate a domestic environment~\cite{turpault_2019,serizel_2020}. The recorded soundscapes are taken from AudioSet~\cite{Gemmeke2017audioset}. The synthetic soundscapes are generated using Scaper~\cite{salamon2017scaper}, a python library for soundscape synthesis and augmentation. The foreground events are obtained from the Freesound Dataset (FSD50k)~\cite{font2013freesound,fonseca2020fsd50k}. The background textures are obtained from the SINS dataset (activity class ``other'')~\cite{Dekkers2017} and TUT scenes 2016 development dataset~\cite{mesaros_tut_2016}.

The dataset includes a validation set and a public evaluation set composed of recorded clips (VAL and EVAL) that are used to adjust the hyper-parameters and evaluate the SED, respectively. All the experiments reported in Tables~\ref{tab:train_set}-\ref{tab:rampup} are performed on VAL.

In order to monitor the SED model convergence, we further split the DESED training set into an training set and a cross-validation set (refereed to as x-valid in the tables). In the public DCASE 2020 task 4 baseline, the cross-validation set was composed of 10\% on the total amount of weakly labeled soundscapes and 10\% of the total amount of synthetic soundscapes generated in the training set\footnote{\url{https://zenodo.org/record/3745475}}.

The difference here is that the cross-validation synthetic soundscapes are generated using a separate set different isolated foreground events that are different from those used to generate the training set (isolated events are split into 90\%/10\% for the train/cross-validation respectively).

\subsection{Sound event detection baseline}
The input features for the SED baseline\footnote{ \url{https://github.com/turpaultn/dcase20_task4/tree/public_branch/baseline}} are mel-spectrograms with 128 mel bands. The signals are sampled at 16~kHz, the mel-spectrogram features are obtained from Short-term Fourier transform coefficient (STFT) computed on 2048 sample windows with 255 hop size.

The CNN block is composed of 7 layers with [16,  32,  64,  128,  128, 128, 128] filters per layer, respectively. We use a kernel size of 3x3 and the max-pooling is [[2, 2], [2, 2], [1, 2], [1, 2], [1, 2], [1, 2], [1, 2]] per layer, respectively. The convolution operations are followed by gated liner unit activation.

The RNN block is composed of 2 layers of 128 bidirectional gated recurrent units. The RNN block is followed by an attention pooling layer that is multiplication between a linear layer with softmax activation and linear layer with sigmoid activation.

The model is trained with Adam optimizer, we apply 50~\% dropout. The model is trained for 200 epochs and the best epoch on the cross-validation set is kept.

When they are used, the SED detection thresholds are fixed to 0.5 for every classes. The post-processing is a median filtering on $\approx$0.45~s (27 frames at 16~kHz).

\subsection{Evaluation metrics}

SED systems are evaluated according to an event-based F1-score with a 200~ms collar on the onsets and a collar on the offsets that is the greatest of 200~ms and 20\% of the sound event's length. The overall F1-score is the unweighted average of the class-wise F1-scores (macro-average). F1-scores are computed on a single operating point (decision thresholds=0.5) using the sed\_eval library~\cite{mesaros_metrics_2016}.

SED systems are also evaluated with poly-phonic sound event detection scores(PSDS)~\cite{Bilen2020}. PSDS are computed using 50 operating points (linearly distributed from 0.01 to 0.99) with the following parameters: detection tolerance parameter ($\rho_{\mathrm{DTC}}=0.5$), ground truth intersection parameter ($\rho_{\mathrm{GTC}} = 0.5$), cross-trigger tolerance parameter ($\rho_{\mathrm{CTTC}} = 0.3$), maximum False Positive rate ($e_{\mathrm{max}} = 100$). The weight on the cost trigger cost is set to $\alpha_{\mathrm{CT}}=1$ and the weight on the class instability cost is set to $\alpha_{\mathrm{ST}}=0$.

\section{Experiments}

The SED baseline has been built on previous year's baseline~\cite{turpault_2019} and challenge submissions~\cite{Lu2018,Delphin-Poulat2019}. As a consequence from this, the SED baseline results from the combination of several technical solutions. The impact of each of these solution has barely been investigated until now. In this part we propose a detailed analysis of the SED baseline system. We propose to study the following aspects:
\begin{itemize}
    \item the type of data used for training;
    \item the transformations applied to the isolated sound events and to the synthetic soundscapes while generating the soundscapes;
    \item the amount of noise added to the mean-teacher input;
    \item the use of ramp-up to balance between the classification loss and the consistency loss;
    \item the use of ramp-up on the learning rate.
\end{itemize}

In the subsequent tables, the column where the performance is 34.14\% corresponds to the original SED baseline setup, trained with the new training/cross-validation split (see also Section~\ref{subsec:desed}). For simplicity sake we report here only the largest confidence intervals obtained during the experiments to provide insight about the significance of the performance difference. For F1-score, performance difference below 1.2\% on the VAL set and below 1.3\% on the EVAL set are generally not to be considered as statistically significant. For the PSDS the intervals are 0.015 and 0.018 on the VAL and EVAL sets, respectively.

One of the key challenge in the task is the way the heterogeneous training data composition is handled. In a first set of experiments we analyse the impact of each subset (recorded clips without labels or weakly labeled, strongly labeled synthetic soundscapes) on the SED performance (Table~\ref{tab:train_set}). We did not include the case when using only the unlabeled data as the SED baseline requires some supervision.

When using only weakly labeled recorded soundscapes, the performance degrades severely. This is due to the lack of strong labels that provide information on how to perform the sound event segmentation at test time. When using only synthetic soundscapes with strong labels, the performance degrades also. In this case, the degradation is probably caused by the acoustic mismatch between the training data (synthetic soundscapes) and the test data (recorded soundscapes).

When combining several subsets, the ratio of soundscapes used per subset has been optimized during preliminary experiments and we present here only the performance obtained with the optimal ratio. When combining only two subset, the combinations including unlabeled data fail to improve the performance. The only combination of two datasets that improves the performance is the combination of the weakly labeled subset and the synthetic soundscapes subset. This combination probably allows for overcoming the weak labels problem and the domain mismatch problem mentioned above. Finally, the best performance is obtained by combining all the subsets (as in the original SED baseline). The diversity provided by the unlabeled subset seems to be beneficial to the SED system.

The second experiment aims at investigating the impact of the transformation used while generating the synthetic soundscapes. The number of isolated sound events in DESED is limited. To overcome this problem, Scaper offers the possibility to apply pitch shifting on the isolated sound events before including them in the soundscapes.

Table~\ref{tab:pitch_synth} presents the SED performance depending on pitch-shifting being applied on the isolated sound events before generating the soundscapes or not. The best combination in terms of F1-score is not to apply pitch-shifting on the training set but to apply it on the cross-validation set. In terms of PSDS the best combination is to apply pitch-shifting on both the training set and the validation (as in the original baseline). This indicates that even if the impact of the pitch shifting on the SED performance is not always significant, using pitch-shifting is beneficial for the SED systems as it increases the diversity of the isolated sound events.

In Scaper there is the possibility to add reverberation to the soundscapes in order to blend isolated sound events in and to avoid having just a juxtaposition of isolated sound events. By default, the reverberation is added with Sox\footnote{http://sox.sourceforge.net/}. In order to have a more realistic reverberation applied to the synthetic soundscapes we are replacing the Sox-based reverberation applied on the soundscapes by room impulse responses (RIR) from the FUSS dataset~\cite{fuss} applied on each isolated sound events.

Table~\ref{tab:reverb_synth} presents the SED performance depending on if reverberation is applied on the synthetic soundscapes or not. The original baseline was not using FUSS reverberation. Applying RIR from FUSS on the validation set improves the performance, possibly because the synthetic soundscapes are then more realistic. Surprisingly, when applying RIR from FUSS on both the training set and the validation set, the performance degrades severely. This can be due to the fact that the RIR in FUSS do not match to the acoustic condition observed in the recorded soundscapes.

The next set of experiments is related to the SED model. In the mean-teacher model, Gaussian white noise is added to the original soundscapes before feeding them to the mean-teacher branch. This is supposed to improve the robustness of the SED model. In the original SED baseline, the SNR between the soundscape and the Gaussian noise was 30 dB. In Table~\ref{tab:SNR} we present the SED performance depending on the SNR applied on the mean-teacher branch. Performance quickly degrades when the SNR decreases. Performing SED at 0dB SNR is really challenging~\cite{serizel_2020}, this could explain why feeding 0dB soundscapes to the mean-teacher branch is actually degrading performance. On the other hand, totally removing the noise from the teacher branch allows for improved performance. An explanation for this could be that the different dropout in the branches is already adding noise between the models.

Optimizing the mean-teacher model involves the combination of several losses (two classification losses and two consistency costs). One frequent problem when optimizing several losses at once is that of balancing between the losses. A solution that have proven to be effective in the case of the mean-teacher is the use of the so-called ramp-up where the weight attributed to one loss is gradually increased across time. Delphin-Poulat et al. proposed to apply the same approach to increase gradually the learning rate (LR)~\cite{Delphin-Poulat2019}. This is commonly called LR warm-up~\cite{goyal2017accurate}.

Table~\ref{tab:rampup} presents the SED performance depending on the weight applied on the consistency costs and depending on whether ramp-up is applied on not. Applying ramp-up on both the consistency costs and the LR (as in the baseline) allows for obtaining the best performance. Additionally, increasing the weight on the consistency cost from 1 to 2 also improves the performance.

Experiments on the SNR in the mean-teacher branch, on pitch-shifting and on reverberation showed that there is room for improvement compared to the original baseline. In table~\ref{tab:reverb_pithc_synth} we propose to perform an additional set of experiments on pitch-shifting and reverberation in the case where there is no additive noise in the mean-teacher branch. In this experiment, we also provide the performance on DESED public evaluation set (EVAL). Here pitch-shifting and reverberation are applied only on the cross-validation set. The best performance (on the VAL set) is obtained when applying pitch-shifting but not reverberation.

\section{Conclusions}
In this paper we provided an detailed analysis of several technical aspects implemented in DCASE task 4 baseline since 2018 (as well as in other SED systems). Some of the default settings that are passed from one system to another without being questioned were shown to be sub-optimal for the task at hand. Through the detailed ablation study, this paper provides insights on how to configure a SED system to be trained on a heterogeneous dataset. The resulting SED system, even-though really similar to DCASE 2020 task 4 official baseline, actually outperforms this baseline by up to 3\% on both the validation and set the public evaluation set.

\section{Acknowledgements}
We would like to thank all the other organizers of DCASE 2020 task 4\footnote{http://dcase.community/challenge2020/task-sound-event-detection-and-separation-in-domestic-environments}. In particular, we would like to thank Justin Salamon and Prem Seetharaman for their help with Scaper and Hakan Erdogan, John R. Hershey and Scott Wisdom for their help with the FUSS dataset.

\bibliographystyle{IEEEtran}
\bibliography{refs}

\end{sloppy}
\end{document}